\documentclass[MSNbibl,number,dvips]{arxstspdf}
\usepackage{flushend}
\usepackage{stfloats}

\volume{26}
\issue{3}
\pubyear{2011}
\firstpage{319}
\lastpage{321}
\doi{10.1214/11-STS352B}
\referstodoi{10.1214/11-STS352}

\begin{document}
\begin{frontmatter}

\vspace*{6pt}
\title{Discussion of ``Is Bayes Posterior just Quick and Dirty Confidence?'' by D. A. S. Fraser}
\runtitle{Discussion}
\pdftitle{Discussion of Is Bayes Posterior just Quick and Dirty Confidence? by D. A. S. Fraser}

\begin{aug}
\author{\fnms{Kesar} \snm{Singh}\corref{}\ead[label=e1]{kesar@stat.rutgers.edu}}
\and
\author{\fnms{Minge} \snm{Xie}\ead[label=e2]{mxie@stat.rutgers.edu}}
\runauthor{K. Singh and M. Xie}

\affiliation{Rutgers University}

\address{Kesar Singh is Professor of Statistics, Department of Statistics and Biostatistics,
Rutgers University, Piscataway, New Jersey 08854, USA \printead{e1}. Minge Xie
is Professor of Statistics, Department of Statistics and Biostatistics,
Rutgers University, Piscataway, New Jersey 08854, USA \printead{e2}.}

\end{aug}

% ABSTRACT

% KEYWORDS

\vspace*{-3pt}
\end{frontmatter}

We congratulate Professor Fraser for this very engaging article. It
gives us an opportunity to gaze at the past and future of Bayes and
confidence. It is well known that a Bayes posterior can only provide
credible intervals and has no assurance of frequentist coverage (known
as confidence). Professor Fraser's article provides a detailed and
insightful exploration into the root of this issue. It turns out that
the Bayes posterior is exactly a confidence in the linear case
(a~mathematics coincidence), and Professor Fraser's insightful and
far-reaching examples demonstrate how the departure from linearity
induces the departure of a posterior, in a proportionate way, from being
a confidence. Of course, Bayesian inference is not bounded by
frequestist criteria or geared to provide confidence statements, even
though in some applications researchers have treated the Bayes credible
intervals as confidence intervals on asymptotic grounds. It is debatable
whe\-ther this departure of Bayesian inference from confidence should be a
concern or not. But, nevertheless, the article provides us a powerful
exploration and demonstration which can help us better comprehend the
two statistical philosophies and the 250-year debate between Bayesians
and frequentists.

%t1 ###
\begin{table*}
\caption{An analogy between ``confidence distribution versus fiducial
distribution'' and ``consistent and asymptotically efficient point
estimator versus MLE''}
\label{tab1}
\begin{tabular*}{\textwidth}{@{\extracolsep{\fill}}p{100pt}p{190pt}p{190pt}@{}}
\hline
% Row 1
& \multicolumn{1}{c}{\textbf{CD (``distribution estimation'')}} & \multicolumn{1}{c@{}}{\textbf{Analogy in point estimation}}\\
\hline
% Row 2
CD definition \& Definition of point estimator & Any sample dependent
distribution function on the parameter space can in principle be used
to estimate the parameter, but we impose a certain requirement (i.e.,
as a function of the sample, the CD is $\operatorname{Uniform}[0,1]$-distributed at the
true parameter value; see, e.g., \cite{6} and \cite{8}) to ensure that the
statistical inferences (e.g., point estimates, confidence intervals,
$p$-values, etc.) derived from the CD have the desired frequentist
property. & Any single point (a real value or a statistic) on the
parameter space can in principle be used to estimate a parameter, but
we impose restrictions so that the point estimator can have certain
desired properties, such as unbiasedness, consistency, efficiency, etc.
\\[50pt]
% Row 3
CD {versus} Fiducial distribution \& Consistent and
asymptotically efficient estimator {versus} MLE & Under some
suitable conditions, fiducial distributions satisfy the frequentist
coverage property (see, e.g., \cite{3} and \cite{4}), which typically make them CD
functions. Thus, the fiducial approach can provide a standard procedure
to obtain a~CD function. & Under some regularity conditions, the MLEs
typically have certain desired frequentist properties (e.g.,
consistency, asymptotic efficiency, etc.). Thus, the MLE approach
provides a~standard procedure to obtain desirable point estimators.\\[3pt]
% Row 4
& A CD does not have to be a fiducial distribution or involve any
fiducial reasoning. & A point estimator with desirable properties does
not have to be an MLE.\\
\hline
\end{tabular*}
\vspace*{6pt}
\end{table*}

In the midst of the 250-year debate, Fisher's ``fiducial distribution''
played a prominent role, which, however, is now referred to as the
``biggest blunder'' of the father of modern statistical
inference~\cite{1}.\vadjust{\goodbreak}
Both developments of the confidence distribution and Fisher's fiducial
distribution share the common goal of providing distribution estimation
for parame\-ters without using priors, and their performances~are often
judged by the (asymptotic or exact) probability coverage of their
corresponding intervals. May\-be partly due to this reason and partly due
to Fisher's ``feud'' and ``longtime dispute'' with Neyman (cf. \cite{13}),
the confidence distribution has historically often been misconstrued as
a fiducial concept. On page 10, Professor Fraser states that ``In the
frequentist framework, the function $p(\theta)$ can be viewed as a
distribution of confidence, as introduced by Fisher \cite{3} but originally
called fiducial.''
It seems to suggest that the concept of
confidence distribution is exchangeable with Fisher's fiducial
distribution. But recent resurging interest and research on confidence
distributions calls for a disagreement with this more classical
assertion. We would like to take the opportunity to raise this point for
discussion. Professor Fraser has a more in-depth understanding of the
issue and he may well wish to correct us, if we are mistaken or have
missed something.

First of all, in the recent developments on confidence distributions,
the concept is developed strictly within the frequentist domain and
resides entirely within the frequentist logic, and there is no
involvement of any fiducial reasoning; see, for example, \cite{6,8,9}.
This can in fact also be seen in all of Professor Fraser's illustrative
examples in the article, in which no fiducial argument is adopted. To
us, a confidence distribution, which uses a sample dependent
distribution on the parameter space to estimate the parameter of
interest, is no different from a point estimator, which uses a (sample
dependent) point in the parameter space to estimate the parameter of
interest. Neither is it different from a confidence interval, which uses
two sample dependent points in the parameter space to estimate the
parameter of interest. In this interpretation, a confidence distribution
is no longer viewed as an inherent distribution of the fixed (nonrandom)
parameter $\theta$ and, unlike the fiducial distribution, it is a
probability distribution in a~frequentist sense. The nice thing about
treating a confidence distribution as a purely frequentist concept is
that the confidence distribution is now a clean and coherent frequestist
concept (similar to a point estimator) and it frees itself from those
restrictive, if not controversial, constraints set forth by Fisher on
fiducial distributions. Table \ref{tab1} uses an analogy to describe the relation
between the new concept of confidence distribution and the fiducial
distribution. A similar analogy was also described in \cite{12} and \cite{10}.

The concept of confidence distribution has attrac\-ted a surge of renewed
attention in recent years. The renewed interest in confidence
distributions starts with Efron \cite{2}, who asserted that bootstrap
distributions are ``distribution estimators'' and ``confidence
distributions.'' He predicted that ``something like fiducial inference''
may ``become a big hit in the 21st century.'' The goal of these new
developments is not to derive any new fiducial inference that is paradox
free. Rather, it is to provide useful statistical inference tools for
problems where frequentist methods with good properties were previously
unavailable or hard to obtain. It seems relevant, without going into
details of specific examples, to indicate a~variety of recent studies
involving confidence distributions, ranging from confidence distribution
and its inference, approximate \mbox{likelihood} \mbox{inference}, incorporation of
expert opinions in a frequesntist setting, combination of information
from independent studies, confidence curves, to applications in survival
analysis and others. We refer interested readers to \cite{2,6,7,8,12,11}
and also a review article~\cite{10}.

As pointed out by Professor Fraser, confidence distribution is a very
old concept first suggested by Neyman in 1937 \cite{5} and some similar ideas
can be traced back even earlier to Bayes \cite{1} and Fisher~\cite{3}. A nagging
question that comes to our mind is why is this concept largely unknown
in the statistical community and why have statisticians never regarded
it as a valuable tool? We believe that inference based on confidence
distributions deserves a~place in the statistician's toolbox and that
distributional inference by confidence distributions should be more
wi\-dely known and used. Many recent research activities on the topic are
aimed at achieving just that. As for Bayesian analysis, confidence or
not, the impact on sciences of this seemingly modest discovery of Thomas
Bayes, in terms of updating (revising) information in light of new data
evidence, is nothing short of miraculous. With the advent of Bayesian
learning, its future couldn't be brighter. Let us be grateful to
Professor Fraser for giving us this opportunity to revisit and examine
the past and future of Bayes and confidence.

\section*{Acknowledgments}
This research is partly supported by NSF Grants DMS-11-07012, DMS-09-15139, SES-08-51521
and NSA Grant H98230-08-1-0104.

% imsref loaded by akundreckaite, 2011-08-04 12:36:52

\end{document}